\documentclass[sn-basic,iicol]{sn-jnl}

\usepackage{graphicx}%
\usepackage{multirow}%
\usepackage{amsmath,amssymb,amsfonts}%
\usepackage{amsthm}%
\usepackage{mathrsfs}%
\usepackage[title]{appendix}%
\usepackage{xcolor}%
\usepackage{textcomp}%
\usepackage{manyfoot}%
\usepackage{booktabs}%
\usepackage{algorithm}%
\usepackage{algorithmicx}%
\usepackage{algpseudocode}%
\usepackage{listings}%
\usepackage{layouts} 
\usepackage{subcaption}
\usepackage{siunitx}
\usepackage{bm} 			

\usepackage{tikz}
\usetikzlibrary{arrows,chains,positioning,quotes,shapes}

\begin{document}

\title{Shape Optimization of Eigenfrequencies in MEMS Gyroscopes}

\author*[1,2]{\fnm{Daniel} \sur{Schiwietz}}\email{daniel.schiwietz@de.bosch.com}
\author[1]{\fnm{Marian} \sur{H\"orsting}}\email{marian.hoersting@de.bosch.com}
\author[2]{\fnm{Eva Maria} \sur{Weig}}\email{eva.weig@tum.de}
\author[1]{\fnm{Peter} \sur{Degenfeld-Schonburg}}\email{peter.degenfeld-schonburg@de.bosch.com}
\author[1]{\fnm{Matthias} \sur{Wenzel}}\email{matthias.wenzel2@de.bosch.com}

\affil*[1]{\orgdiv{Corporate Sector Research and Advance Engineering}, \orgname{Robert Bosch GmbH}, \orgaddress{\city{Renningen}, \postcode{71272}, \country{Germany}}}

\affil[2]{\orgdiv{School of Computation, Information and Technology}, \orgname{Technical University of Munich}, \orgaddress{\city{Munich}, \postcode{80333}, \country{Germany}}}

\abstract{Microelectromechanical systems (MEMS) gyroscopes are widely used in consumer and automotive applications. They have to fulfill a vast number of product requirements which lead to complex mechanical designs of the resonating structure. Arriving at a final design is a cumbersome process that relies heavily on human experience in conjunction with design optimization methods. In this work, we apply node-based shape optimization to the design of a MEMS gyroscope. For that purpose, we parametrize the coordinates of the nodes of the finite element method (FEM) mesh that discretize the shapes of the springs. We then implement the gradients of the mechanical eigenfrequencies and typical MEMS manufacturability constraints, with respect to the design parameters, in a FEM code. Using gradient-based optimization we tune the gyroscope's frequency split and shift spurious modes away from the first three multiples of the gyroscope's drive frequency while manufacturability constraints are fulfilled. The resulting optimized design exhibits novel geometrical shapes which defy any human intuition. Overall, we demonstrate that shape optimization can not only solve optimization problems in MEMS design without required human intervention, but also explores geometry solutions which can otherwise not be addressed. In this way, node-based shape optimization opens up a much larger space of possible design solutions, which is crucial for facing the ever increasing product requirements. Our approach is generic and applicable to many other types of MEMS resonators.}


\maketitle

\section{Introduction}\label{sec1}
Microelectromechanical systems (MEMS) resonators are omnipresent in everyday life. They enable a broad range of sensors, actuators and filters in automotive, consumer and communication applications \citep{neul_micromachined_2007,shaeffer_mems_2013,wang_review_2021,pillai_piezoelectric_2021}. Gyroscopes, i.e., angular rate sensors, are a particularly complex application which is found in almost every modern smartphone and car. In such use cases, the sensor has to remain operable under various harsh external conditions \citep{acar_environmentally_2009,minin_mems_2011}. At the same time, the requirements to the sensor's performance and miniaturization are increasing with every generation. This leads to very complex mechanical designs of gyroscopes which are usually the result of an elaborate design process. This design process relies heavily on manual topology adjustments and human intuition \citep{benkhelifa_design_2010}. As a result, the final designs are typically an intricate combination of simple building blocks, usually of rectangular shape, with well known scaling laws for properties such as eigenfrequencies \citep{giannini_size_2020}.

Structural optimization can help human designers arrive faster at a final design or even obtain designs which are beyond human intuition. Typically, structural optimization is divided into three distinct approaches, which are size optimization, shape optimization and topology optimization \citep{christensen_introduction_2008}. In size optimization, the design variables are structural thicknesses or lengths, for example of beams. Size optimization provides the smallest design freedom and has the least amount of design variables for a given structure. In shape optimization, the boundary of structural elements, such as beams, is optimized. In a discretized geometry, the design variables control the positions of the bounding nodes of the structure. As a result, shape optimization provides much more design freedom than size optimization and has a larger amount of design variables, due to the possibly large amount of bounding nodes. However, it still retains the connectivity of structural elements. Finally, topology optimization provides the largest design freedom. Here, the design variables typically define whether a cell of a discretized geometry is occupied by material or not. As a result, the connectivity between structural elements can change.

All three structural optimization approaches have been explored in the field of MEMS resonators. The majority of publications has focused on size optimization. In this case, typical design parameters in MEMS resonators are lengths and widths of springs and beams. Both gradient-based \citep{giannini_size_2020,dou2015structural,li2017tailoring} and gradient-free methods \citep{xia_design_2015,kacprzyk_evolutionary_2010,shin2022spiderweb} have been explored. Size optimization enables only small design variations with respect to the initial design. Therefore, basic functionality can be ensured by providing a good initial design which is then fine-tuned by the optimization algorithm. Furthermore, the small amount of design variables allows the use of gradient-free methods. However, due to the limited number of design variables, complex optimization problems might not be solvable. Alternatively, topology optimization has been applied to obtain unintuitive designs with tailored eigenfrequencies \citep{giannini_topology_2020,giannini_topology_2022,he_topology_2012} and quality factors \citep{hoj2021ultra,gao2020systematic}. However, MEMS resonators often have to fulfill a vast amount of requirements. In order to obtain feasible designs via topology optimization, the gradients of all requirements would have to be implemented, which can be prohibitively complex. Shape optimization can be seen as a compromise between size and topology optimization \citep{christensen_introduction_2008}. Here, the initial topology is maintained, ensuring basic functionality. At the same time, the design space is much larger than in size optimization and designs which are beyond human intuition can be obtained. Shape optimization is a well-known method, that has been applied to a variety of problems in other fields \citep{hojjat_vertex_2014,ghantasala_realization_2021,antonau_latest_2022}.

Various shape parametrization methods can be applied for the purpose of optimization \citep{samareh2001survey}. Free-form deformation embeds the geometry in a fictitious flexible material, defined by a set of lattice points. The deformation of the lattice is then propagated to the geometry. By that, the geometry change instead of the geometry is parametrized, allowing to control a complex geometry via a small set of design variables \citep{kenway2010cad}. Another approach is to use B-splines, which are commonly employed in computer-aided design (CAD), to define surfaces through a set of control points. The control points can be taken as design variables for optimization \citep{lee2017comparison}. Alternatively, the level set method combines aspects of shape and topology optimization, by describing the surface implicitly through a level set function. The boundary of the shape evolves through the level set function and splitting and merging of boundaries is possible \citep{luo2008level}. A node-based approach is given by vertex morphing. Here, each design variable corresponds to the location of one surface node's location and a filter is applied to obtain smooth designs \citep{hojjat_vertex_2014}. Node-based shape optimization, compared to other shape parametrization methods, provides the highest possible number of design variables and can be readily applied to any design given by a finite element method (FEM) model.

However, to our knowledge, node-based shape optimization has not been applied to MEMS resonators, yet. We believe that node-based shape parametrization is well suited for MEMS design, as FEM models of the geometry are available anyways during the design process. In this work, we apply node-based shape optimization to the design of a MEMS gyroscope to tailor its eigenfrequencies while maintaing manufacturability constraints.

In section~\ref{sec:shapeoptimization} we introduce the MEMS gyroscope and its basic operating principle. Furthermore, we provide the theoretical background of the optimization procedure. For that purpose, we explain the definition of the design parameters, derive relevant manufacturability constraints, formulate the optimization problem and detail the optimization loop. In section~\ref{sec:results}, we provide the results of the MEMS gyroscope's shape optimization and discuss them. In section~\ref{sec:conclusion}, we summarize our work and conclude that shape optimization is a very promising approach for MEMS resonator design.

\section{Shape Optimization of MEMS Gyroscopes} \label{sec:shapeoptimization}
In this section we lay out the theoretical background relevant to the shape optimization of MEMS resonators. Furthermore, we introduce the MEMS gyroscope on which we will demonstrate the shape optimization methodology. Our approach is applicable to any kind of MEMS resonator, particularly useful for extruded geometries that can easily be manufactured via etching. For this work, we chose MEMS gyroscopes, as they constitute some of the most complex MEMS resonators, with a wide range of commercial applications. After giving an overview over the MEMS gyroscope, along with its working principle, design and boundary conditions, we detail the definition of relevant design parameters. Furthermore, we will pay particular attention to manufacturability constraints, which are customary for MEMS resonators. Finally, we will formulate the shape optimization problem and summarize the relevant steps of the optimization loop.

\subsection{Modeling of the MEMS Gyroscope}
\begin{figure*}
	\centering
	\includegraphics[width=0.9\textwidth]{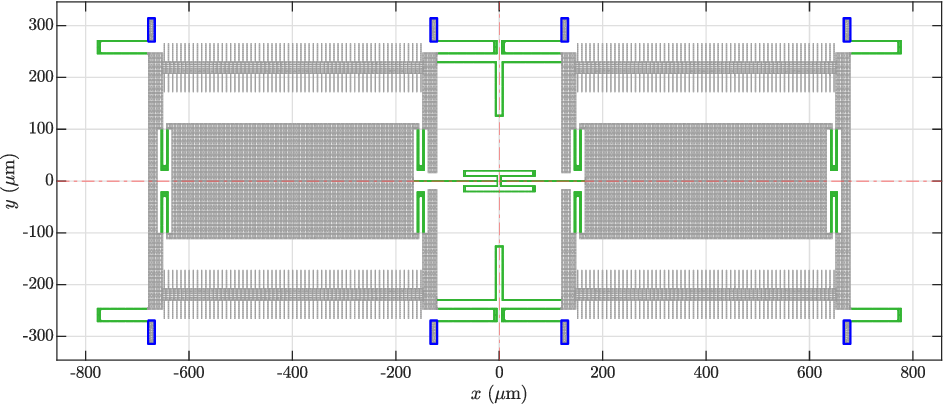}
	\caption{Top view of the single-axis MEMS gyroscope's initial design. The red lines indicate the symmetry axes, the substrate anchors are outlined in blue and the springs are colored in green.}
	\label{fig:InitialDesign}
\end{figure*}
\begin{figure*}
	\begin{subfigure}{0.5\textwidth}
		\centering
		\includegraphics[width=\textwidth]{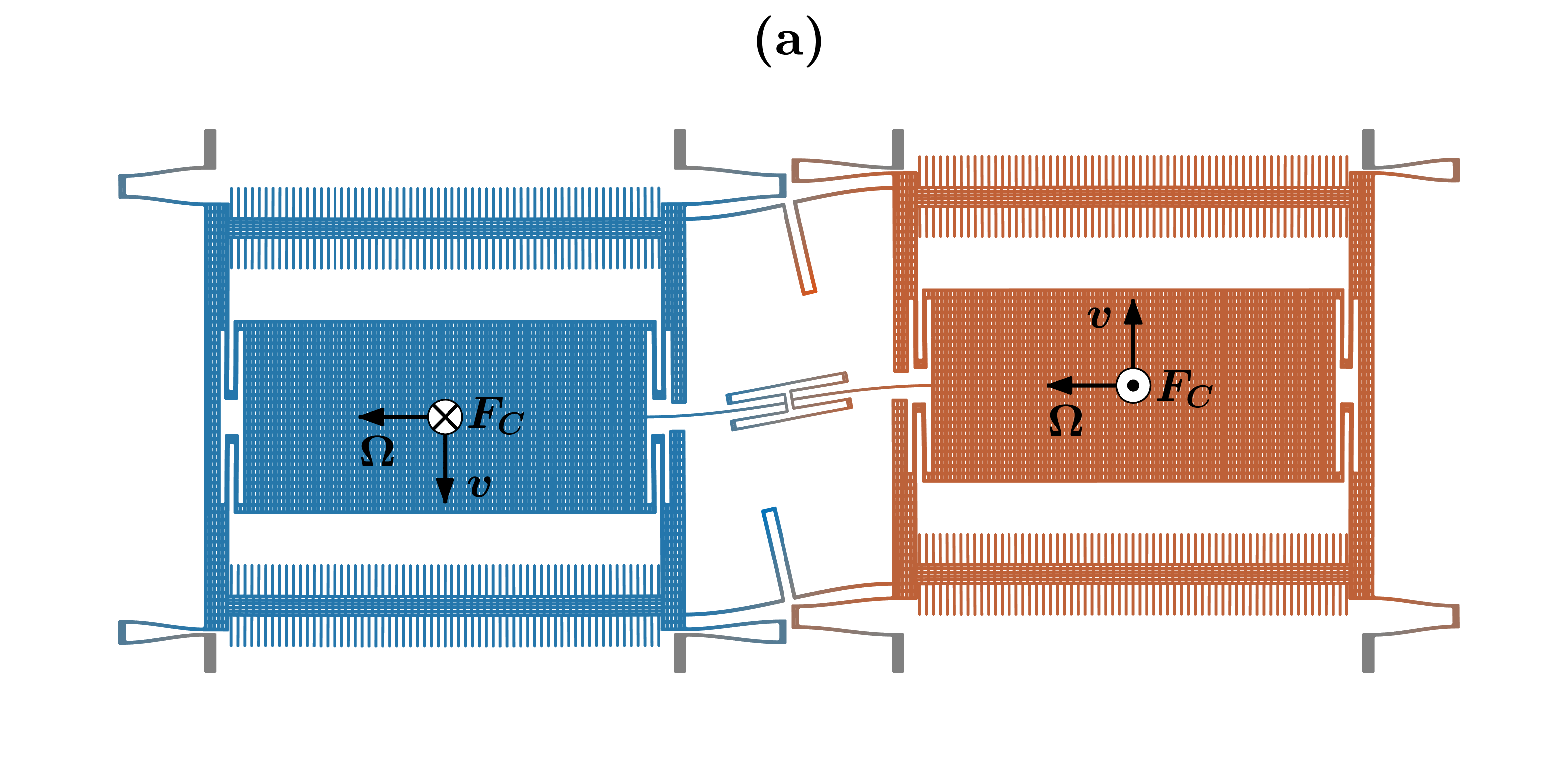}
	\end{subfigure}
	\hfill
	\begin{subfigure}{0.5\textwidth}
		\centering
		\includegraphics[width=\textwidth]{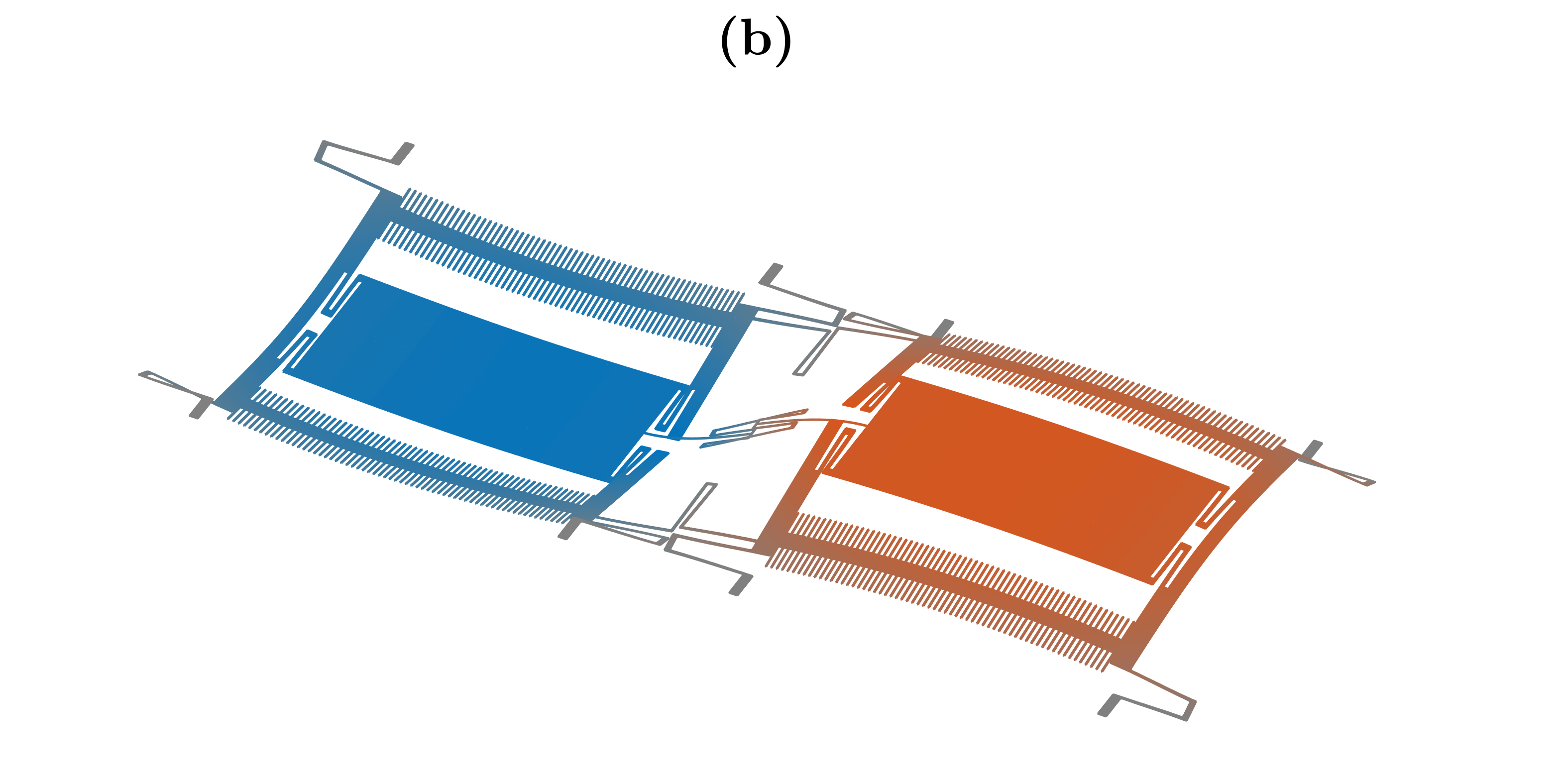}
	\end{subfigure}
	\caption{Mode shapes of the initial design. Only the displacement of the uppermost nodes is shown for simplicity. The coloring indicates the sum of the three displacement components at each point (red: positive, blue: negative). (a): Drive mode with an eigenfrequency of \SI{24.4}{\kilo\hertz}. Indicated is also the Coriolis force acting on each mass as a result of its velocity and the applied angular rate. (b): Detection mode with an eigenfrequency of \SI{27.3}{\kilo\hertz}.}
	\label{fig:ModeShapes}
\end{figure*}
We will employ shape optimization to shift eigenfrequencies of a single-axis MEMS gyroscope. The same MEMS gyroscope has already been subject of previous publications, which dealt with the modeling of geometric nonlinearities \citep{putnik_simulation_2016,putnik_incorporating_2017,putnik_static_2017,putnik_predicting_2018,putnik_simulation_2018}. In this work, we focus on the linear eigenfrequencies.

A top view of the gyroscope's initial design is shown in Fig.~\ref{fig:InitialDesign}. The material is assumed to be linear isotropic polycrystalline silicon with Poisson's ratio $\nu=\SI{0.2261}{}$, Young's modulus $E=\SI{161}{\giga\pascal}$ and density $\rho=\SI{2330}{\kilogram\per\cubic\meter}$. The functional layer of MEMS gyroscopes is typically grown via epitaxy and referred to as the expitaxial layer. The out-of-plane thickness of the epitaxial layer is \SI{10.55}{\micro\metre}. The design exhibits a quarter symmetry and has two substrate anchors per quadrant. The substrate anchors are the connections of the functional layer to the underlying substrate, which is not included in the model. The anchoring is modeled by applying fixed boundary conditions underneath the anchors. The anchors are outlined in blue in Fig.~\ref{fig:InitialDesign}. The gyroscope contains two masses, which are connected to each other and to the anchors via springs. The springs are colored in green in Fig.~\ref{fig:InitialDesign}. Furthermore, each mass holds two sets of comb electrodes. Figure~\ref{fig:ModeShapes} shows the drive mode, which oscillates in-plane with an eigenfrequency of \SI{24.4}{\kilo\hertz}, and the detection mode, which oscillates out-of-plane with an eigenfrequency of \SI{27.3}{\kilo\hertz}, for the initial design. During operation, the drive mode is forced to oscillate harmonically by means of an alternating voltage applied to the comb electrodes. When subjected to an angular rate $\bm{\Omega}$ around the x-axis, combined with the velocity $\bm{v}$ due to the oscillation along y-direction, the structure experiences a Coriolis force $\bm{F}_C$, which excites the detection mode. This is indicated in Figure~\ref{fig:ModeShapes}~(a). The angular rate can then be measured from the differential capacitance signal of the out-of-plane electrodes underneath the two masses.

Due to the quarter symmetry of the design, we perform all simulations on a quarter model. The full set of eigenmodes is obtained by applying the four different combinations of symmetric and anti-symmetric boundary conditions on the two symmetry axes. The quarter model is discretized by around 49,000 quadratic hexahedral elements with 20 nodes per element and a total of around 330,000 nodes. The mesh was created in two dimensions (2D) and then extruded along the out-of-plane direction with 3 elements along the epitaxial thickness. Therefore, the obtained mesh consists of nodes that are perfectly aligned above each other along the z-direction - a characteristic property which we will exploit during the shape optimization. After applying the appropriate boundary conditions, the eigenvectors, or mode shapes, and eigenfrequencies of the quarter model are obtained from a modal analysis which solves the generalized eigenvalue problem
\begin{equation}
	\left(\bm{K}-\omega_i^2\bm{M}\right)\bm{\phi}_i=\bm{0}, \label{eq:gevp}
\end{equation}
with stiffness matrix $\bm{K}$, mass matrix $\bm{M}$, eigenvector $\bm{\phi}_i$, angular eigenfrequency $\omega_i=2\pi f_i$ and eigenfrequency $f_i$ of mode $i$. The eigenvectors are mass-normalized, i.e., $\bm{\phi}_i^T\bm{M}\bm{\phi}_i=1$. The full model's eigenvectors can be obtained by mirroring the eigenvectors of the quarter model appropriately.

\subsection{Definition of Design Parameters} \label{sec:designparameters}
\begin{figure}
	\centering
	\includegraphics[width=0.45\textwidth]{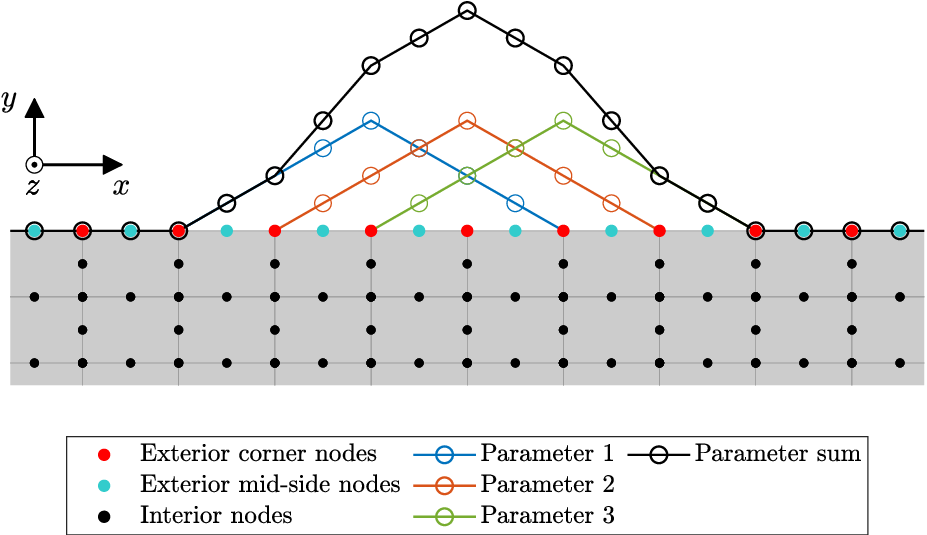}
	\caption{Definition of the design parameters. Each parameter corresponds to a hat function, that is centered on an exterior corner node. This leads to a smoothing of the optimized shape, due to neighboring nodes being dragged along. The sum of the applied shifts of all parameters yields the boundary of the optimized shape. For illustration purposes, only three design parameters are shown.}
	\label{fig:HatFunctions}
\end{figure}
\begin{figure}
	\centering
	\begin{subfigure}{0.15\textwidth}
		\centering
		\includegraphics[width=\textwidth]{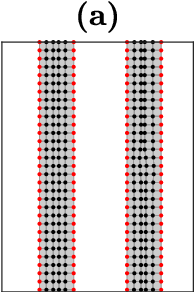}
	\end{subfigure}
	\hfill
	\begin{subfigure}{0.15\textwidth}
		\centering
		\includegraphics[width=\textwidth]{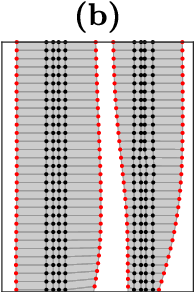}
	\end{subfigure}
	\hfill
	\begin{subfigure}{0.15\textwidth}
		\centering
		\includegraphics[width=\textwidth]{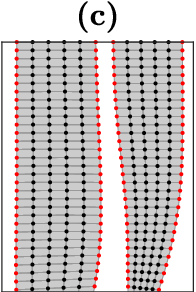}
	\end{subfigure}
	\\
	\vspace{0.1cm}
	\centering
	\begin{subfigure}{0.45\textwidth}
		\centering
		\includegraphics[width=\textwidth]{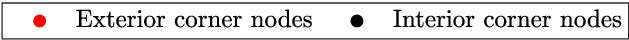}
	\end{subfigure}
	\caption{Update of the corner nodes. The mid-side nodes are afterwards placed centrally between neighboring corner nodes. (a):~Initial mesh. (b):~Displacement of the exterior corner nodes. (c):~Update of the interior corner nodes. }
	\label{fig:MeshUpdate}
\end{figure}
The dynamic properties of the gyroscope, such as its eigenfrequencies, are determined by the shape of the structure. Consequently, in a discretized FEM model, the positions of the boundary nodes, also referred to as exterior nodes, determine the shape and thus the physical properties. The interior nodes only have an indirect influence, as a low quality interior mesh can worsen the approximation error of the FEM simulation. This motivates shape optimization based on mesh morphing. The geometry of the initial design, as shown in Fig.~\ref{fig:InitialDesign}, is essentially an arrangement of rectangular shapes, as is customary for MEMS gyroscopes. Shifting the boundary nodes allows one to obtain complex and unintuitive shapes, which can influence the gyroscope's properties in a desirable way.

As the eigenfrequencies are mainly determined by the springs, we will parametrize the positions of their exterior nodes. This means that the shape of the green regions in Fig.~\ref{fig:InitialDesign} will be morphed. We define the design parameters such that they shift the exterior nodes along the surface normal in the xy-plane. For that purpose, we project the geometry onto the xy-plane, by only considering the topmost layer of nodes, which we will refer to as 2D nodes. In our model, the mid-side nodes of the employed serendipity elements are always centered between neighboring corner nodes of the same element. Thus, we only assign design parameters to corner nodes, as they determine the geometry. Therefore, each 2D exterior corner node, belonging to a spring, is assigned a design parameter. 

To avoid sharp kinks in the optimized design, we employ vertex morphing by introducing hat functions to smoothen the nodal shifts \citep{hojjat_vertex_2014}. On each node, that corresponds to a design parameter, a hat function is centered and neighboring exterior 2D nodes are dragged along, according to the hat function's value. The hat functions are chosen such that they decay over a range of $d_{max}=\SI{15}{\micro\metre}$, traced along the surface in the initial design. This is shown exemplarily in Fig.~\ref{fig:HatFunctions}. The coordinates of the mesh are therefore parametrized as
\begin{equation}
\bm{x}=\bm{x}_0+\frac{\partial\bm{x}}{\partial\bm{p}}\bm{p}, \label{eq:dxdp}
\end{equation}
where $\bm{x}$ is a vector which contains the $x-$, $y$- and $z$-coordinates of all nodes, $\bm{x}_0$ contains the positions of the nodes for the initial design and $\bm{p}$ is a vector containing the design parameter values. The surface normals in the xy-plane are calculated at every exterior node in the initial design. The direction, along which the nodes are shifted by a parameter, is then determined by the surface normal of the node at the center of the hat function. The surface normals are not recalculated throughout the optimization. Therefore, $\frac{\partial\bm{x}}{\partial\bm{p}}$ is constant during the optimization. We normalize the assignment of our design parameters such that the sum of displacement magnitudes of all 2D nodes affected by parameter $p_j$ is equal to \SI{1}{\micro\metre} for a parameter value of $p_j=1$. Thus, the design parameters affect the 2D nodes as
\begin{equation}
\frac{\partial\bm{x}_i}{\partial p_j}=\frac{\bm{n}_{j,0}\max(d_{max}-\lvert d_{i,j,0} \rvert,0)}{\sum_k\max(d_{max}-\lvert d_{k,j,0} \rvert,0)},
\end{equation}
where $\bm{x}_i$ are the coordinates of node $i$, $\bm{n}_{j,0}$ is the surface normal of node $j$ in the initial design and is normalized as $\left\Vert\bm{n}_{j,0}\right\Vert=1$, $d_{i,j,0}$ is the distance along the surface between nodes $i$ and $j$ in the initial design and the index $k$ runs over all exterior 2D nodes. The nodes which are not part of the 2D mesh have the same $\frac{\partial\bm{x}_i}{\partial p_j}$ assigned as their corresponding 2D nodes at the same xy-location. Therefore, all nodes that are aligned above each other in z-direction are shifted equally by a given design parameter. In total we define 3413 parameters for the optimization.

Furthermore, we also update the position of the interior nodes of the mesh, after each design update. This is done to ensure a high mesh quality and is done via a static 2D FEM analysis \citep{yao_3d_1989}. We consider the 2D projection of the initial mesh in the xy-plane. The displacements of the exterior corner nodes, with respect to their initial positions, as defined by $\frac{\partial\bm{x}}{\partial\bm{p}}\bm{p}$, are applied as Dirichlet boundary conditions. The displacements of the interior corner nodes, with respect to the initial mesh, are then obtained as the solution of a 2D linear static analysis. However, due to the large displacements of the exterior nodes, this sometimes leads to invalid elements. Therefore, we subsequently minimize the distortion of the resulting 2D mesh, with respect to the positions of the interior corner nodes, similar to \cite{ruiz-girones_simultaneous_2015}. The mid-side nodes are then simply centered between neighboring corner nodes. Finally, all interior nodes, which are stacked above each other along the z-direction, are then shifted by the values determined from the 2D mesh. The update of exterior and interior corner nodes is illustrated in Fig.~\ref{fig:MeshUpdate}.

\subsection{Manufacturability Constraints}
\begin{figure}
	\centering
	\includegraphics[width=0.45\textwidth]{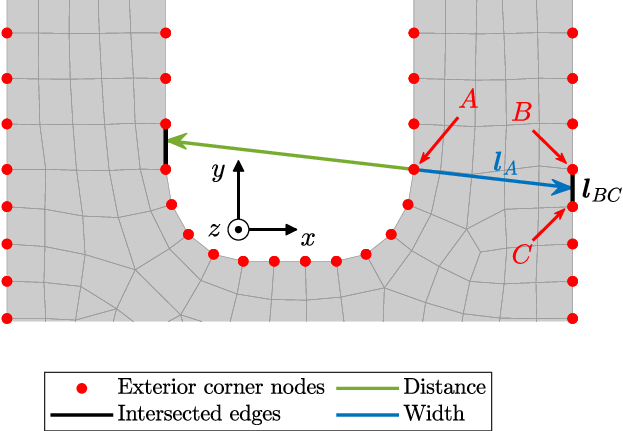}
	\caption{Visualization of the distance and width constraints for a single exterior node $A$. Starting from the exterior node, the arrows follow the surface normal in positive and negative direction until they intersect another exterior element edge. The lengths of the arrows then determine the distance and width at the exterior node. The relevant lines and points are labeled for the width constraint.}
	\label{fig:DistanceWidthConstraint}
\end{figure}
Typical constraints in MEMS manufacturing are the minimum allowable width of features and minimum allowable distance between features, such as mass and spring building blocks. In order to obtain designs, which are manufacturable, these constraints have to be included in the optimization.

We determine the width and distance at an exterior node by tracing the surface normal in positive and negative direction, until it intersects another exterior element edge. The length of the traced path then determines the value of the constraint. The sign of the path, i.e., whether the path went inward or outward with respect to the surface, determines whether it is a width or distance constraint. This is visualized for a single exterior node in Fig.~\ref{fig:DistanceWidthConstraint}. Distance and width constraints are introduced for every exterior node which has a design parameter centered on it. Note that, since we are only concerned with the distances and widths in the xy-plane, the manufacturability constraints can be entirely calculated in 2D. As with the definition of the design parameters, we again always use the surface normals of the initial design for the calculations. We found that, for the given MEMS gyroscope optimization, this yields sufficient fulfillment of the manufacturability constraints.

Starting from an exterior node $A$, with 2D position vector $\bm{x}_A$ and 2D surface normal in the initial design $\bm{n}_{A,0}$, the equation describing the line $\bm{l}_A$, that is perpendicular to the surface at node $A$, reads
\begin{equation}
\bm{l}_A=\bm{x}_A + s\bm{n}_{A,0}, \label{eq:l_A}
\end{equation}
where $s$ is a parameter that quantifies the position along the line. Similarly, for each element with an exterior boundary, we can describe the edge, which is cornered by exterior corner nodes $B$ and $C$, by a line $\bm{l}_{BC}$ as
\begin{equation}
\bm{l}_{BC}=\bm{x}_B + t(\bm{x}_C - \bm{x}_B), \label{eq:l_BC}
\end{equation}
where $\bm{x}_B$ and $\bm{x}_C$ are the 2D position vectors of nodes $B$ and $C$ and $t$ is a parameter that quantifies the position along the line. The intercept of the two lines is found by equating Eqs.~\eqref{eq:l_A} and \eqref{eq:l_BC}, leading to
\begin{equation}
\begin{bmatrix}
s\\
t
\end{bmatrix} 
=
\begin{bmatrix}
\bm{n}_{A,0}, & (\bm{x}_B-\bm{x}_C)
\end{bmatrix}^{-1}
(\bm{x}_B-\bm{x}_A), \label{eq:dist}
\end{equation}
which can be easily evaluated for many combinations of lines, as the inverted matrix is only of size 2 by 2. We normalize the surface normals as $\lVert \bm{n}_{A,0} \lVert=1$. The surface normal intersects the exterior edge, cornered by nodes $B$ and $C$, if $0 \leq t \leq 1$. The closest intersection along the positive direction of the surface normal defines the distance constraint and the closest intersection along the negative direction defines the width constraint for node $A$. The value of the constraint is given by the magnitude of $s$. The relevant points and lines are labeled in Fig.~\ref{fig:DistanceWidthConstraint} for an exemplary width constraint. A kd-tree is used to vastly reduce the number of line segments which have to be checked for a given node $A$. Kd-trees are data structures that partition points in space and are commonly used for ray tracing. They can be employed to efficiently find close geometric entities for which an intersection is then checked \citep{hapala2011kd}.

\subsection{Optimization Problem Formulation}
The eigenfrequencies of the mechanical modes are decisive for the functionality of MEMS resonators, in particular MEMS gyroscopes. Tailoring the eigenfrequencies is an essential part during the design of MEMS gyroscopes, as will be explained below. Several approaches have been presented in literature, in order to aid and automate the process of obtaining a design with specified target eigenfrequencies \citep{giannini_size_2020,giannini_topology_2020,giannini_topology_2022,he_topology_2012,xia_design_2015,kacprzyk_evolutionary_2010}. Shape optimization has two significant advantages compared to previously employed methods. Firstly, in contrast to size optimization, the design space increases tremendously, due to the large amount of exterior nodes that exist in complex MEMS structures. Secondly, basic sensor functionality is maintained, since the initial topology is preserved, which is an advantage compared to topology optimization.

We will demonstrate the potential of shape optimization in MEMS gyroscope design by optimizing the eigenfrequencies of the MEMS gyroscope shown in Fig.~\ref{fig:InitialDesign}. As the goal of the optimization, we wish to obtain a certain split $\Delta f=f_{\text{det}}-f_{\text{drv}}$, between the detection mode's eigenfrequency $f_{\text{det}}$ and the drive mode's eigenfrequency $f_{\text{drv}}$, while keeping $f_{\text{drv}}$ fixed within $\pm1\%$ of its initial value $f_{\text{drv},0}$. At the same time, we wish to shift the eigenfrequencies of all spurious modes, i.e., all modes that aren't detection or drive, away from the $\pm10\%$ regions around $1f_{\text{drv}}$, $2f_{\text{drv}}$ and $3f_{\text{drv}}$. Furthermore, we minimize the applied design change, i.e., we search for a design which is close to the initial design.

The optimization goals are motivated by typical requirements during MEMS gyroscope design. The out-of-plane detection mode's eigenfrequency can be tuned by electrostatic softening via an out-of-plane plate capacitor. The in-plane drive mode is actuated by comb electrodes, which exhibit no electrostatic softening due to their constant gap \citep{acar_mems_2009}. Therefore, a positive $\Delta f$ allows to obtain mode-matched operation despite process tolerances. The $\Delta f$ should exceed the process-induced frequency variations, without being so large that impractically high voltages are required for the electrostatic softening. For this work, we target a $\Delta f$ between \SI{1.9}{\kilo\hertz} and \SI{2.1}{\kilo\hertz}, which represents a typical pratical range.

Furthermore, shifting the eigenfrequencies of the spurious modes away from  $1f_{\text{drv}}$, $2f_{\text{drv}}$ and $3f_{\text{drv}}$ helps to avoid unwanted nonlinear phenomena, such as internal resonances, which can deteriorate the device's performance. Spurious modes at multiples of the drive mode's eigenfrequency can be excited by square wave signals typically used to drive MEMS gyroscopes, through geometrically nonlinear couplings with the drive mode or through nonlinear electrostatic forces \citep{nabholz_nonlinear_2019,younis2011mems,gobat2021reduced,wagner2021determining}.

Mathematically the optimization problem is written as
\begin{equation}
\begin{aligned}
\min_{\bm{p}}       & \quad  \bm{p}\cdot\bm{p}         \\
\text{subject to: } & \quad 0.99f_{\text{drv},0} \leq f_{\text{drv}} \leq 1.01f_{\text{drv},0} \\
& \quad \SI{1900}{\hertz} \leq \Delta f \leq \SI{2100}{\hertz} \\
& \quad H_{n,i} \leq 0  \\ 
& \quad n=1,2,3 \\
& \quad i=1,\ldots,n_m \quad i\neq\text{drv},\text{det} \\
& \quad d_j\geq d_{min} \\ 
& \quad w_j\geq w_{min} \\ 
& \quad -9 \leq p_j \leq 9 \\
& \quad j=1,\ldots,n_p \label{eq:OptimizationProblem}
\end{aligned}
\end{equation}
with the function $H_{n,i}$, which is positive when mode $i$ is within $\pm10\%$ of $nf_{\text{drv}}$, and the total number of modes $n_m=36$. The manufacturability constraints include the distance $d_j$ and width $w_j$ constraints. We choose, exemplarily, a minimum distance of $d_{min}=\SI{2}{\micro\meter}$ and a minimum width of $w_{min}=\SI{1.5}{\micro\meter}$. The number of design parameters, i.e., the dimension of $\bm{p}$, is denoted by $n_p=3413$. The minimization in Eq.~\eqref{eq:OptimizationProblem} ensures that the design is close to the initial design; the constraint on $f_{\text{drv}}$ fixes the frequency of the drive mode; the constraint on $\Delta f$ determines the frequency split; the constraints on $H_{n,i}$, which are formulated for all spurious modes, shift the eigenfrequencies out of the $\pm10\%$ regions around $1f_{\text{drv}}$, $2f_{\text{drv}}$ and $3f_{\text{drv}}$; the constraints on $d_j$ and $w_j$ ensure the manufacturability. The limits on the design parameter values $p_j$ were chosen such that a converged solution could be obtained without being unnecessarily large. We define the function $H_{n,i}$ as
\begin{equation}
	H_{n,i} = 1-\frac{|f_i-nf_{\text{drv}}|}{0.1nf_{\text{drv}}}. \label{eq:Hni}
\end{equation}
Equation~\eqref{eq:Hni} has a value of 1 when $f_i$ is at $nf_{\text{drv}}$ and decays linearly to zero over a range of $0.1nf_{\text{drv}}$. Beyond this range the function becomes negative. With respect to convergence, we found this continuous formulation to be superior compared to a formulation that is set to zero outside of the allowed region.

\subsection{Optimization Algorithm}
\begin{figure*}
	\centering
	\begin{tikzpicture}
	\tikzstyle{startstop} = [rectangle, draw, text centered, rounded corners, minimum width=3cm, minimum height=2em, fill=red!30]
	\tikzstyle{block} = [rectangle, draw, text centered, minimum width=10cm, minimum height=2em, fill=orange!30]
	\tikzstyle{decision} = [diamond, minimum width=6cm, minimum height=1cm, text centered, draw, fill=green!30, aspect=6]
	\tikzstyle{arrow} = [thick,->,>=stealth]
	
	\node (initial) [startstop] {Initial design};
	\node (parameters) [block, below of=initial, yshift=-0.25cm] {Definition of design parameters};
	\node (modana) [block, below of=parameters, yshift=-0.25cm] {Modal analysis and mode tracking};
	\node (constraints) [block, below of=modana, yshift=-0.25cm] {Calculate objective function, constraints and sensitivities};
	\node (convergence) [decision, below of=constraints, yshift=-0.5cm] {All constraints satisfied?};
	\node (mma) [block, below of=convergence, yshift=-0.5cm] {Determine new positions of exterior nodes using MMA};
	\node (update) [block, below of=mma, yshift=-0.25cm] {Update interior node positions based on 2D static analysis};
	\node (final) [startstop, below of=update, yshift=-0.25cm] {Final design};
	
	\draw [arrow] (initial) -- (parameters);
	\draw [arrow] (parameters) -- (modana);
	\draw [arrow] (modana) -- (constraints);
	\draw [arrow] (constraints) -- (convergence);
	\draw [arrow] (convergence) to ["No"] ++ (0,-1) -- (mma);
	\draw [arrow] (mma) -- (update);
	
	\draw [arrow] (update.west) to ++ (-0.5,0) |- (modana);
	
	\draw [arrow] (convergence.east) to ["Yes"] ++ (2,0) |- (final);
	\end{tikzpicture}
	\caption{Flow chart of the shape optimization algorithm.}
	\label{fig:AlgorithmFlowChart}
\end{figure*}
In this section, the details of the shape optimization algorithm will be discussed. The general steps of the algorithm are summarized in Fig.~\ref{fig:AlgorithmFlowChart}.

At the beginning, the meshed initial geometry has to be provided. We created the initial design and its mesh in ANSYS. The initial mesh, i.e., the node coordinates and element connectivity, was then imported into MATLAB. From this point on, the entire algorithm is implemented in a self-written vectorized MATLAB code. All FEM calculations, including the assembly of system matrices, the calculation of sensitivities and the modal analysis, are performed within the code.

Based on the initial mesh, the nodes belonging to the springs, as shown in Fig.~\ref{fig:InitialDesign}, are identified. Design parameters are then defined, based on the exterior nodes of the springs, as described in section~\ref{sec:designparameters}. By that, $\frac{\partial\bm{x}}{\partial\bm{p}}$ is determined and remains constant throughout the optimization. The goal of the algorithm is to find the values of $\bm{p}$ which lead to the satisfaction of all constraints in Eq.~\eqref{eq:OptimizationProblem} while minimizing the change of design variables.

Next, a modal analysis is performed for the first 36 modes, to ensure that all modes that could be in the vicinity of $3f_{\text{drv}}$ are included. During the optimization, the order of the modes might change. To have each constraint consistently refer to the same modes throughout all iterations, we track the modes as described in Appendix~\ref{sec:A_modetracking}.

After the modal analysis has been performed and the modes have been identified, the constraints in Eq.~\eqref{eq:OptimizationProblem} can be evaluated based on the current geometry and eigenfrequencies. Note that, in this work, we don't consider any constraints that depend on eigenvectors. Furthermore, as we employ gradient-based optimization, the gradients, or sensitivities, of the objective function and all constraints with respect to the design parameters are required. All gradients are calculated analytically, based on the defined $\frac{\partial\bm{x}}{\partial\bm{p}}$. Most importantly, the sensitivities of the eigenfrequencies with respect to the design parameters are required.

Taking the derivative of Eq.~\eqref{eq:gevp} with respect to parameter $p_j$ and then taking the dot product with eigenvector $\bm{\phi}_i$, one finds the well-known result \citep{tortorelli_design_1994} for the sensitivity of the $i$-th mode's eigenfrequency $f_i$ as
\begin{equation}
\frac{\partial f_i}{\partial p_j}=\frac{1}{4\pi\omega_i}\bm{\phi}_i^T\left(\frac{\partial \bm{K}}{\partial p_j}-\omega_i^2\frac{\partial \bm{M}}{\partial p_j}\right)\bm{\phi}_i, \label{eq:eigenvaluesensitivity}
\end{equation}
which requires the sensitivities of the FEM system matrices $\frac{\partial \bm{K}}{\partial p_j}$ and $\frac{\partial \bm{M}}{\partial p_j}$. Note that the products in Eq.~\eqref{eq:eigenvaluesensitivity} can be evaluated element-wise and then summed up. Therefore, the sensitivities of the element matrices only have to be calculated for elements which have a non-zero sensitivity for a given parameter. This enables an efficient calculation. The analytic calculation of $\frac{\partial \bm{K}}{\partial p_j}$ and $\frac{\partial \bm{M}}{\partial p_j}$ is obtained based on the initially defined $\frac{\partial\bm{x}}{\partial\bm{p}}$ and has to be recalculated in each design iteration, since the sensitivities of the system matrices are also functions of the nodal coordinates. Further details on the calculation of the sensitivies of the optimization problem in Eq.~\eqref{eq:OptimizationProblem} are given in appendix~\ref{sec:A_derivatives}.

The optimization is terminated when all constraints in Eq.~\eqref{eq:OptimizationProblem} are fulfilled. The objective function serves to guide the optimization to a final design which is close to the initial design, as there might be multiple designs which fulfill all constraints.

If the current design doesn't meet the optimization goal, the optimization problem in Eq.~\eqref{eq:OptimizationProblem} and its sensitivities with respect to all design parameters are passed to a gradient-based optimizer, which provides a new $\bm{p}$. For this purpose, we chose the method of moving asymptotes (MMA) \citep{svanberg_method_1987}.

Based on the new $\bm{p}$, the positions of the exterior nodes can be updated according to Eq.~\eqref{eq:dxdp}. Furthermore, to maintain a high mesh quality, the positions of the interior mesh's nodes are also updated, as described in section~\ref{sec:designparameters}.

The resulting mesh is then evaluated in the next iteration. The steps are iterated until the optimization goal is met.

Note that the gradient calculation is based on $\frac{\partial\bm{x}}{\partial\bm{p}}$, which does not include the shift of the interior nodal positions. In other words, $\frac{\partial\bm{x}}{\partial\bm{p}}$ has zeros at all entries that correspond to interior nodes. We found that the indirect influence of parameters on interior nodes has negligible impact on the sensitivities of Eq.~\eqref{eq:OptimizationProblem} and can be neglected.

\section{Results}\label{sec:results}
\begin{figure*}
	\centering
	\includegraphics[width=0.9\textwidth]{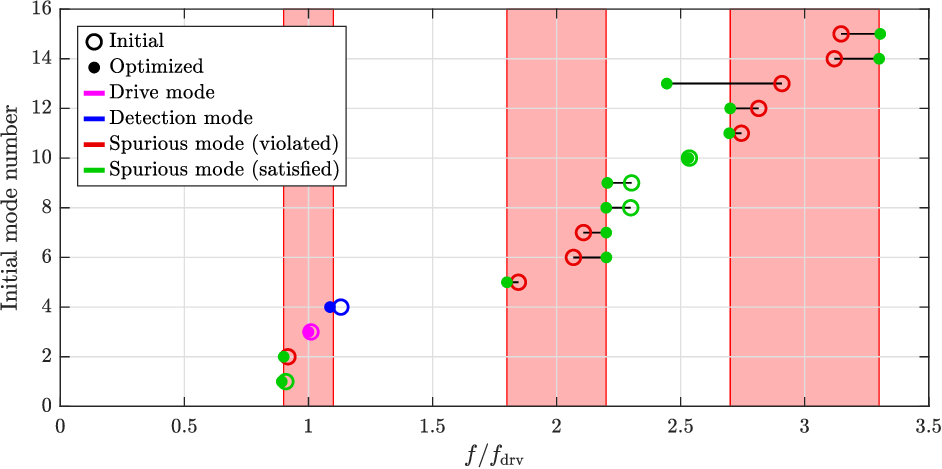}
	\caption{Initial and optimized frequency spectrum. The eigenfrequencies of the modes are normalized to the value of the drive mode's eigenfrequency in the optimized design. Empty circles show the spectrum of the initial design and filled dots show the spectrum of the optimized design. The black lines connect the modes from initial and optimized design that correspond to each other according to the mode tracking. Green markers indicate spurious modes that are outside of the forbidden frequency bands, whereas red markers indicate spurious modes that are inside of the forbidden frequency bands. The color-coding of the empty circles refers to their positions with respect to the frequency bands in the initial design. The red areas indicate the forbidden frequency bands in the optimized design.}
	\label{fig:ResultsSpectrum}
\end{figure*}
\begin{figure*}
	\centering
	\includegraphics[width=0.9\textwidth]{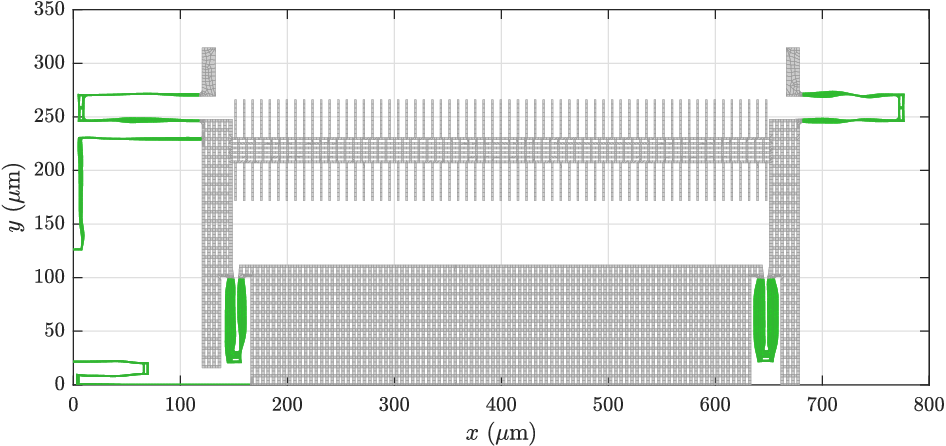}
	\caption{Final quarter model of the converged shape optimization. The full design has a quarter symmetry, which is exploited in the optimization procedure. The springs are colored in green.}
	\label{fig:OptimizedQuarter}
\end{figure*}
\begin{figure*}
	\centering
	\begin{subfigure}{0.45\textwidth}
		\centering
		\includegraphics[height=3.5cm]{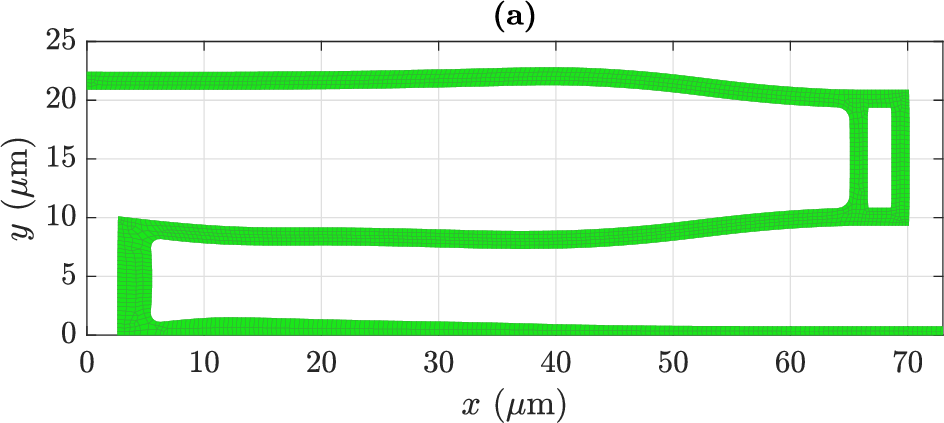}
	\end{subfigure}
	\qquad
	\centering
	\begin{subfigure}{0.45\textwidth}
		\centering
		\includegraphics[height=3.5cm]{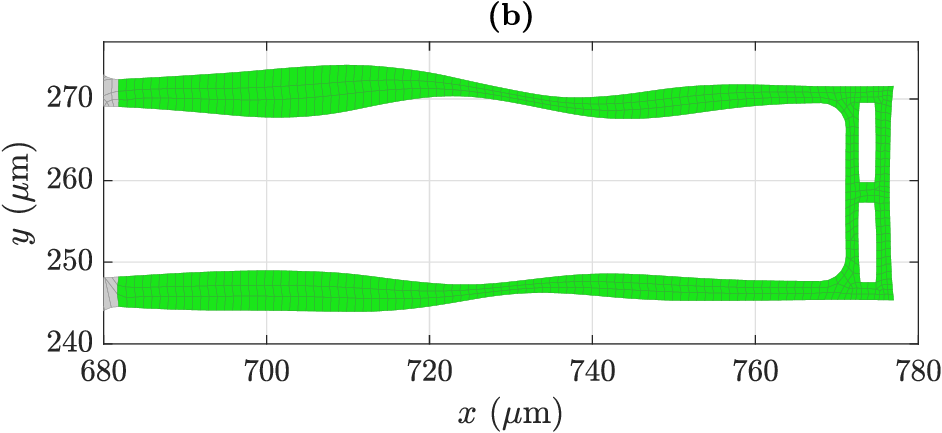}
	\end{subfigure}
	\\
	\vspace{0.1cm}
	\centering
	\begin{subfigure}{0.15\textwidth}
		\centering
		\includegraphics[height=6cm]{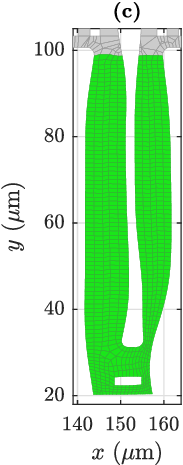}
	\end{subfigure}
	\qquad
	\begin{subfigure}{0.15\textwidth}
		\centering
		\includegraphics[height=6cm]{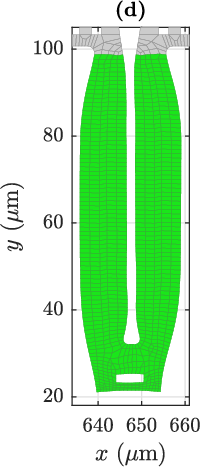}
	\end{subfigure}
	\qquad
	\begin{subfigure}{0.3\textwidth}
		\centering
		\includegraphics[height=6cm]{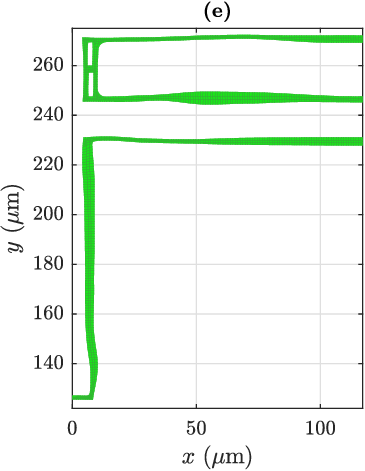}
	\end{subfigure}
	\caption{(a)-(e): Magnified view of the optimized springs in Fig.~\ref{fig:OptimizedQuarter}.}
	\label{fig:OptimizedSprings}
\end{figure*}
Here we discuss the result of the optimization procedure.

Figure~\ref{fig:ResultsSpectrum} compares the frequency spectra obtained on the initial design and on the optimized layout. It took 59 iterations to arrive at the optimized design. The optimized design shows all the desired properties. The drive mode's eigenfrequency decreased slightly, but less than 1\%. The detection mode's eigenfrequency shifted to lower values in order to decrease the frequency split $\Delta f$ from \SI{2.9}{\kilo\hertz} to \SI{2.1}{\kilo\hertz}. The spectrum of the initial design had 9 spurious modes within the prohibited frequency bands, as shown by the red empty circles in Fig.~\ref{fig:ResultsSpectrum}. In the optimized design, the eigenfrequencies of all spurious modes are outside of these frequency bands, as shown by the filled green dots in Fig.~\ref{fig:ResultsSpectrum}. Furthermore, all width and distance constraints have been fulfilled.

Due to the sensor's symmetry, the optimization was performed on a quarter model of the sensor, corresponding to the first quadrant in Fig.~\ref{fig:InitialDesign}. This approach reduces computation time and maintains perfect quarter symmetry. It was conducted on a high-performance computer with 12 CPU cores, taking around 7 hours and using a maximum of 110 GB of memory.

The optimized spring layout is shown in Fig.~\ref{fig:OptimizedQuarter}. All springs of the optimized design show significant morphing. A magnified view of the optimized springs is provided in Fig.~\ref{fig:OptimizedSprings}. The two springs connecting the central mass to the frame have become significantly wider, as seen in Figs.~\ref{fig:OptimizedSprings}~(c) and (d). The widest parts of the springs are around \SI{11}{\micro\meter} wide. To obtain a movable structure, the functional layer of a gyroscope is typically released from the substrate via etching. Structures being too wide without perforations can lead to the part not being fully released by the etchant. However, we note that this structure is still manufacturable, i.e., underetching would be possible. Note that, due to the way that we calculated the width and distance constraints, it is not straightforward to prescribe maximal values in a universal way. For instance, with our approach, some nodes at the head of a spring would have the entire length of the spring assigned as their width. Looking at the springs on the left in Fig.~\ref{fig:OptimizedQuarter}, also shown in Figs.~\ref{fig:OptimizedSprings}~(a) and (e), which connect the quadrants, we see not just a width variation but also significant shape changes of the initial beam shapes. The two topmost springs in Fig.~\ref{fig:OptimizedQuarter}, also shown in Figs.~\ref{fig:OptimizedSprings}~(b) and (e), which provide the connections to the anchors, exhibit very localized width changes. In total, we find very non-intuitive design changes, which lead to favorable properties and conform with typical design rules for MEMS gyroscopes.

\begin{figure}
	\centering
	\includegraphics[width=0.45\textwidth]{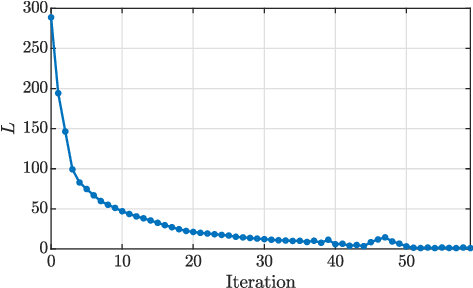}
	\caption{The aggregate function Eq.~\eqref{eq:lagrangian}, summarizing the convergence of objective function and constraints.}
	\label{fig:ConvergenceHistory}
\end{figure}

To illustrate the convergence behavior of the optimization, we write all constraints in the form $\alpha_q c_q\leq0$, where $c_q$ is the value of the $q$-th inequality constraint and $\alpha_q$ is a weight. The weights $\alpha_q$ were chosen as 0.2 for the drive frequency and frequency split constraints, 25 for the spurious mode constraints and 1 for the manufacturability constraints. Additionally, a weight of $\alpha_p=0.1n_p^{-1}$ was chosen for the objective function. By that, the different orders of magnitude of objective function and constraints are accounted for. Note that the same weights were also applied during the optimization. The convergence of objective function and constraints can then be visualized using the function
\begin{equation}
	L = \alpha_p\bm{p}\cdot\bm{p} + \sum_q \max(\alpha_q c_q,0),
	\label{eq:lagrangian}
\end{equation}
which aggregates the objective function and the violated constraints and is shown in Fig.~\ref{fig:ConvergenceHistory}. It can be seen that the optimization converged overall smoothly, besides a few bumps towards the end.

\section{Conclusion}\label{sec:conclusion}
To summarize, we implemented state-of-the-art shape optimization for MEMS resonators in a self-written FEM code introducing constraints to ensure that these can be manufactured with standard MEMS processes. We applied the method to the design of a MEMS gyroscope with typical frequency constraints. The optimized design exhibited springs with very non-intuitive shapes, which visibly deviated from the rectangular structures that are typically found in state of the art MEMS gyroscopes. Our results highlight that shape optimization is a powerful tool to solve complex nonlinear optimization problems that arise during MEMS resonator design. Limiting the geometry to simple shapes necessitates manual topological redesigns which are time consuming. Shape optimization on the other hand can unravel the full potential of a given topology. We believe that MEMS resonators are well suited for shape optimization and that our work will pave the way towards new and more efficient design approaches. Other potential MEMS applications, besides gyroscopes, are accelerometers, micro mirrors, speakers and microphones as well as RF MEMS. In the future we plan to extend our work to more complex constraints and optimization problems, which also depend on the mode shapes of the resonator. Furthermore, experimental validation of shape optimized MEMS structures will be subject of future research. 

\backmatter

\bmhead{Acknowledgements}
The authors are thankful to Martin Putnik at Robert Bosch GmbH for providing the model of the initial design. Furthermore, the authors gratefully acknowledge Daniel Baumg\"artner at Robert Bosch GmbH for the valuable discussions about shape optimization.

\section*{Declarations}

\subsection*{Author contributions}
D.S., P.D. and M.W. crafted the original idea. D.S., M.H. and M.W. implemented the method into code. D.S. performed the optimization. D.S. wrote the manuscript. P.D., E.W. and M.W. supervised the work. All authors reviewed the manuscript.

\subsection*{Funding}
The IPCEI ME/CT project is supported by the Federal Ministry for Economic Affairs and Climate Action on the basis of a decision by the German Parliament, by the Ministry for Economic Affairs, Labor and Tourism of Baden-W\"urttemberg based on a decision of the State Parliament of Baden-W\"urttemberg, the Free State of Saxony on the basis of the budget adopted by the Saxon State Parliament, the Bavarian State Ministry for Economic Affairs, Regional Development and Energy and financed by the European Union - NextGenerationEU.

\subsection*{Conflict of interest}
On behalf of all authors, the corresponding author states that there is no conflict of interest. 

\subsection*{Replication of results}
The manuscript provides all the information needed to replicate the demonstrated method. Furthermore, the model of the MEMS gyroscope can be made available upon reasonable request to the corresponding author.

\begin{appendices}
\section{Mode Tracking}\label{sec:A_modetracking}
We define a modal assurance criterion (MAC) as
\begin{equation}
\text{MAC}^{\{k\}}_{ij} = \left(\bm{R}^{\{k\}}\bm{\phi}^{\{k\}}_i\right)^T\left(\bm{R}^{\{k-1\}}\bm{\phi}^{\{k-1\}}_j\right), \label{eq:MAC}
\end{equation}
where $\bm{R}$ is the upper triangular matrix obtained from the Cholesky decomposition $\bm{M}=\bm{R}^T\bm{R}$, the superscript $\{k\}$ refers to quantities calculated in the current iteration $k$ and $\{k-1\}$ refers to the previous iteration. The mode $i$ for which $\text{MAC}^{\{k\}}_{ij}$ has the largest magnitude determines the mode number in iteration $k$ corresponding to mode $j$ from the previous iteration. We found that the formulation in Eq.~\eqref{eq:MAC} leads to a reliable identification of modes in our optimization.

\section{Sensitivities}\label{sec:A_derivatives}

The sensitivities of Eq.~\eqref{eq:OptimizationProblem} with respect to all design parameters are required for the gradient-based optimization. The sensitivity of the objective function is trivially obtained.

The sensitivities of the frequency-dependent constraints are readily obtained via the chain rule, once the sensitivities of the eigenfrequencies are obtained from Eq.~\eqref{eq:eigenvaluesensitivity}. Evaluating Eq.~\eqref{eq:eigenvaluesensitivity} is the most time consuming step in each optimization iteration. It can be equivalently calculated as a sum over all elements
\begin{equation}
\frac{\partial f_i}{\partial p_j}=\frac{1}{4\pi\omega_i}\sum_{e}(\bm{\phi}^e_i)^T\left(\frac{\partial \bm{K}^e}{\partial p_j}-\omega_i^2\frac{\partial \bm{M}^e}{\partial p_j}\right)\bm{\phi}^e_i, \label{eq:eigenvaluesensitivity_e}
\end{equation}
where $\bm{\phi}^e_i$ contains the components of eigenvector $i$ at the nodes of element $e$ and $\bm{K}^e$ and $\bm{M}^e$ are the stiffness and mass matrix of element $e$. The derivatives $\frac{\partial \bm{K}^e}{\partial p_j}$ and $\frac{\partial \bm{M}^e}{\partial p_j}$ are only non-zero in elements that are morphed by design parameter $p_j$. Therefore, the sensitivities of the element matrices only have to be calculated for the elements that are morphed by a given parameter. For our model, between 50 and 300 elements are morphed per design parameter. This allows the evaluation of Eq.~\eqref{eq:eigenvaluesensitivity_e} for thousands of design parameters and many mechanical modes within reasonable computation time.

\begin{figure}
	\centering
	\includegraphics[width=0.45\textwidth]{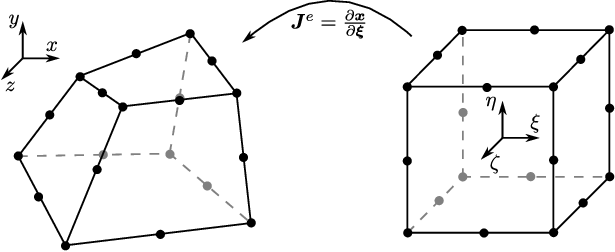}
	\caption{Isoparametric mapping between physical element with coordinates $\bm{x}=(x,y,z)$ and reference element with coordinates $\bm{\xi}=(\xi,\eta,\zeta)$.}
	\label{fig:IsoparametricMapping}
\end{figure}

The sensitivities of the element matrices can be derived from their definitions
\begin{align}
\bm{K}^e&=\sum_p (\bm{B}_\varepsilon^e)^T \bm{D} \bm{B}_\varepsilon^e \det(\bm{J}^e) W_p, \label{eq:kue}\\ 
\bm{M}^e&=\sum_p \rho\bm{N}_u^T\bm{N}_u \det(\bm{J}^e) W_p, \label{eq:me}
\end{align}
where $p$ denotes the integration point index, $\bm{B}_\varepsilon^e$ is a matrix that contains shape function derivatives, $\bm{D}$ is the elasticity matrix, $\bm{J}^e$ is the Jacobian matrix, which describes the transformation into the reference element, as illustrated in Fig.~\ref{fig:IsoparametricMapping}, $W_p$ is the integration point weight, $\rho$ is the density and $\bm{N}_u$ is a matrix containing shape functions. The superscript $e$ denotes quantities that are specific to element $e$. The matrix $\bm{N}_u$ depends on the integration point but is independent of global coordinates. The matrices $\bm{B}_\varepsilon^e$ and $\bm{J}^e$ are functions of the node coordinates and have to be calculated at each integration point. The explicit dependence on node coordinates enters the FEM equations through the Jacobian matrix
\begin{equation}
\bm{J}^e=\bm{x}^e\bm{B}_{\xi}^T, \label{eq:jacobian}
\end{equation}
where $\bm{x}^e$ is a matrix in which each column corresponds to the coordinates of one node belonging to element $e$ and the matrix $\bm{B}_{\xi}$ contains the derivatives of shape functions with respect to the reference element's coordinates. The columns in $\bm{B}_{\xi}$ correspond to the shape functions and the rows correspond to the derivatives with respect to the reference element's coordinates. The derivatives of the shape functions inside element $e$ with respect to the global coordinates are collected in a matrix $\bm{B}^e$, which is obtained from
\begin{equation}
\bm{B}^e=(\bm{J}^e)^{-T}\bm{B}_{\xi}. \label{eq:bxe}
\end{equation}
The columns in $\bm{B}^e$ correspond to the shape functions and the rows correspond to the derivatives with respect to the global coordinates. The matrix $\bm{B}_\varepsilon^e$, which is required in Eq.~\eqref{eq:kue}, is assembled from the entries of $\bm{B}^e$.

Based on the definition of the design parameters, the derivative of the Jacobian is straightforward to obtain from Eq.~\eqref{eq:jacobian} as
\begin{equation}
\frac{\partial \bm{J}^e}{\partial p_j}=\frac{\partial \bm{x}^e}{\partial p_j}\bm{B}_{\xi}^T, \label{eq:djac_dp}
\end{equation}
where $\frac{\partial \bm{x}^e}{\partial p_j}$ is simply a rearrangement of specific entries from $\frac{\partial \bm{x}}{\partial \bm{p}}$. Next, the derivative of Eq.~\eqref{eq:bxe} can be calculated and, after using Eqs.~\eqref{eq:bxe} and~\eqref{eq:djac_dp}, reads
\begin{equation}
\frac{\partial \bm{B}^e}{\partial p_j}=-\bm{B}^e\left(\frac{\partial \bm{x}^e}{\partial p_j}\right)^T\bm{B}^e. \label{eq:dbxe_dp}
\end{equation}
The derivative $\frac{\partial \bm{B}_\varepsilon^e}{\partial p_j}$ is simply assembled from the entries of $\frac{\partial \bm{B}^e}{\partial p_j}$ in the same way that $\bm{B}_\varepsilon^e$ is assembled from the entries of $\bm{B}^e$. Finally, the sensitivity of the Jacobian determinant is required and, by using Eqs.~\eqref{eq:bxe} and~\eqref{eq:djac_dp}, can be written as
\begin{equation}
\frac{\partial \det(\bm{J}^e)}{\partial p_j}=\bm{B}^e:\frac{\partial \bm{x}^e}{\partial p_j}\det(\bm{J}^e). \label{eq:djacdet_dp}
\end{equation}
After having calculated $\frac{\partial \bm{B}_\varepsilon^e}{\partial p_j}$ and $\frac{\partial \det(\bm{J}^e)}{\partial p_j}$, the sensitivities of the element matrices are found by taking the derivatives of Eqs.~\eqref{eq:kue} and~\eqref{eq:me}, leading to
\begin{equation}
\begin{aligned}
\frac{\partial \bm{K}^e}{\partial p_j}&=\sum_p \left[ \left(\frac{\partial \bm{B}_\varepsilon^e}{\partial p_j}\right)^T \bm{D} \bm{B}_\varepsilon^e \det(\bm{J}^e) \right. \\
		&+ (\bm{B}_\varepsilon^e)^T \bm{D} \frac{\partial \bm{B}_\varepsilon^e}{\partial p_j} \det(\bm{J}^e) \\
	    &\left. + (\bm{B}_\varepsilon^e)^T \bm{D} \bm{B}_\varepsilon^e \frac{\partial \det(\bm{J}^e)}{\partial p_j} \right] W_p, \label{eq:dkue_dp}
\end{aligned}
\end{equation}
\begin{equation}
\frac{\partial \bm{M}^e}{\partial p_j}=\sum_p \rho\bm{N}_u^T\bm{N}_u \frac{\partial \det(\bm{J}^e)}{\partial p_j} W_p. \label{eq:dme_dp}
\end{equation}
Equations~\eqref{eq:dkue_dp} and~\eqref{eq:dme_dp} are the exact analytic sensitivities of the element matrices which can also be found in literature \citep{christensen_introduction_2008}. Afterwards, Eq.~\eqref{eq:eigenvaluesensitivity_e} can be evaluated and the sensitivities of eigenfrequency-dependent constraints in Eq.~\eqref{eq:OptimizationProblem} can be obtained from the chain rule. 

Lastly, the sensitivities of the distance and width constraints are needed. They can be obtained from Eq.~\eqref{eq:dist}. Defining
\begin{equation}
\bm{X}:=\begin{bmatrix}
\bm{n}_{A,0}, & (\bm{x}_B-\bm{x}_C)
\end{bmatrix},
\end{equation}
\begin{equation}
\bm{a}:=\bm{x}_B-\bm{x}_A,
\end{equation}
the derivative of Eq.~\eqref{eq:dist} reads
\begin{equation}
\frac{\partial}{\partial p_j}
\begin{bmatrix}
s\\
t
\end{bmatrix} 
=-\bm{X}^{-1}\frac{\partial \bm{X}}{\partial p_j}\bm{X}^{-1}\bm{a} + \bm{X}^{-1}\frac{\partial \bm{a}}{\partial p_j}.
\end{equation}
The derivatives $\frac{\partial \bm{X}}{\partial p_j}$ and $\frac{\partial \bm{a}}{\partial p_j}$ are easily calculated from the entries corresponding to the $x$- and $y$-coordinates of nodes $A$, $B$ and $C$ in $\frac{\partial \bm{x}}{\partial \bm{p}}$. Note that $\frac{\partial \bm{n}_{A,0}}{\partial p_j}=\bm{0}$, since we always use the normal vectors that were defined on the initial design. The derivative $\frac{\partial s}{\partial p_j}$ is the sensitivity of the width for node $A$, when $s$ is negative, and the sensitivity of the distance, when $s$ is positive.

\end{appendices}

\end{document}